# Langevin Dynamics of Chiral Phase Transition at Finite Temperature and Density

Tomoi Koide

*Institute für Theoretische Physik, J.W.Goethe Universität, D-60054, Frankfurt am Main, Germany*

**Abstract.** We derive the linear Langevin equation to describe the dynamics of the chiral phase transition above the critical temperature by applying the projection operator method to the Nambu-Jona-Lasinio model at finite temperature and density. The relaxation time of the critical fluctuations increases as the temperature approaches toward the critical temperature because of the critical slowing down. The critical slowing down is enhanced in low temperature and large chemical potential region and around the tricritical point.

Heavy-ion collisions are essentially nonequilibrium processes and hence the description of the time evolution is necessary to understand the phenomena in a comprehensive way. Then, the hydrodynamic model is a powerful tool, and fairly correctly describes the experimental data of heavy-ion collisions.

However, it is suspicious to use the Hydrodynamic model (in particular, the perfect fluid) to describe dynamics near phase transitions. Actually, the mode coupling theory tells us that we should use Langevin equations instead of the Navier-Stokes equation to describe the critical dynamics[1]. For example, classical fluid usually follows the Navier-Stokes equation. However, near the liquid-gas phase transition, we should solve the following Langevin equations,

$$\frac{\partial}{\partial t}\phi = -\nabla \cdot (\phi \mathbf{v}) + L_0 \nabla^2 \frac{\delta}{\delta \phi} \int d^3\mathbf{x} f(\phi) + \theta, \qquad (1)$$

$$\rho \frac{\partial}{\partial t}\mathbf{v} = -\nabla \cdot \Pi(\phi) + \eta_0 \nabla^2 \mathbf{v} + \zeta, \qquad (2)$$

where $f(\phi)$ and $\Pi(\phi)$ are the Landau free energy and the stress tensor, respectively. The critical dynamics of the liquid-gas phase transition is described remarkably well by this coupled Langevin equation[2].

In this study, we derive the Langevin equation that describes the critical dynamics in chiral phase transition[3]. As a low energy effective model of QCD, we adopt the Nambu-Jona-Lasinio model and apply the projection operator method to derive the Langevin equation[4, 5]. In this method, it is possible to eliminate the irrelevant information with the help of projection operators and obtain simultaneous equations for gross variables. In this case, we choose the chiral condensate $\bar{q}q$ as a gross variable and, for simplicity, we ignore the nonlinear terms and the mode coupling between various gross variables.

In Fig. 1, we show the time dependence of the fluctuations of the chiral condensate $\delta\sigma(\mathbf{0},t) = \sigma(\mathbf{0},t) - \langle \sigma(\mathbf{0},0) \rangle_{eq}$ described by the derived linear Langevin equation at $\mu = 200$ MeV[5]. One can see that the nonequilibrium fluctuations relax exhibiting oscillation and finally converge to zero. This indicates that the critical dynamics of the chiral phase transition may not be described by an overdamping equation like the time-dependent Ginzburg-Landau (TDGL) equation. This means that the assumption employed to derive the TDGL equation may lose its validity in the chiral phase transition. In the derivation of the TDGL equation, one assumes that the time evolution of the order parameter is induced by a sort of the thermodynamic force given by $\delta F/\delta M$, where $F$ is the Ginzburg-Landau free energy and $M$ is the order parameter. However, in general, a number of thermodynamic forces are possible to induce such an irreversible process because of the cross effect. Then, the usual TDGL equation can be changed from an overdamping equation. As another possibility, this oscillating behavior may be related to causality problem in relativistic irreversible systems[6]. On the other hand, the damping rate of the critical fluctuations becomes smaller as the temperature approaches toward the critical temperature. This is because of the critical slowing down, where the long wave length component of the order parameter shows extraordinary large fluctuations near the critical point due to the increase of the relaxation time.

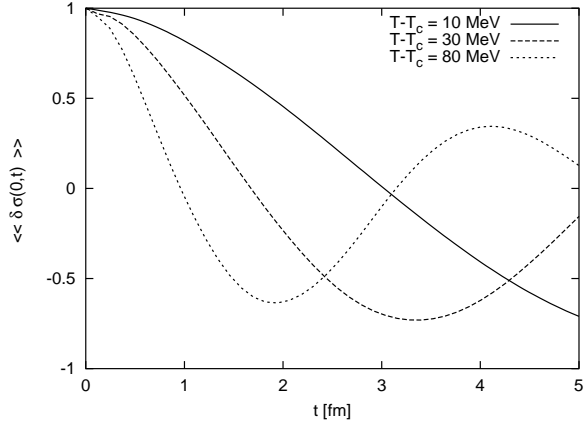

**FIGURE 1.** The time evolution of $\ll \delta\sigma(\mathbf{0},t) \gg$ at $\mu = 200$ MeV. The solid, dashed and dotted lines are the $\ll \delta\sigma(\mathbf{0},t) \gg$ for temperatures $T - T_c = 10$ MeV, 30 MeV, 80 MeV.

In order to discuss the critical slowing down in chiral phase transition, we show the temperature and chemical potential dependences of the relaxation time in Fig. 2[5]. At a fixed $\mu$, the relaxation time increases as the temperature is lowered toward the critical temperature and diverges at the critical point because of the critical slowing down. In this figure, we started to plot from 1 MeV higher temperature than the critical temperature. When we increase $\mu$ along the critical line, we encounter the enhancement of the critical slowing down; one is in the low temperature and large chemical potential region and the other is around the tricritical point. The former can be explained by the Pauli blocking that interferes the decay of the fluctuations into fermions.

The latter would be related to the broadness of the bottom of the thermodynamic potential. When we increase the chemical potential along the critical line, the bottom of the thermodynamic potential increases gradually and achieves maximum at the tricritical point. At larger chemical potential, the thermodynamic potential has two minimum but the potential barrier between the two minima is still shallow. Thus, the large fluctuations are still possible to survive. In short, there exist extraordinary large fluctuations around the tricritical point and this gives rise to the enhancement. Because of the large relaxation time of the critical fluctuations, thermalization is decelerated, in particular, in the low temperature and large chemical potential region and around the tricritical point, although we have ignored the effect of the large sound wave called the piston effect.

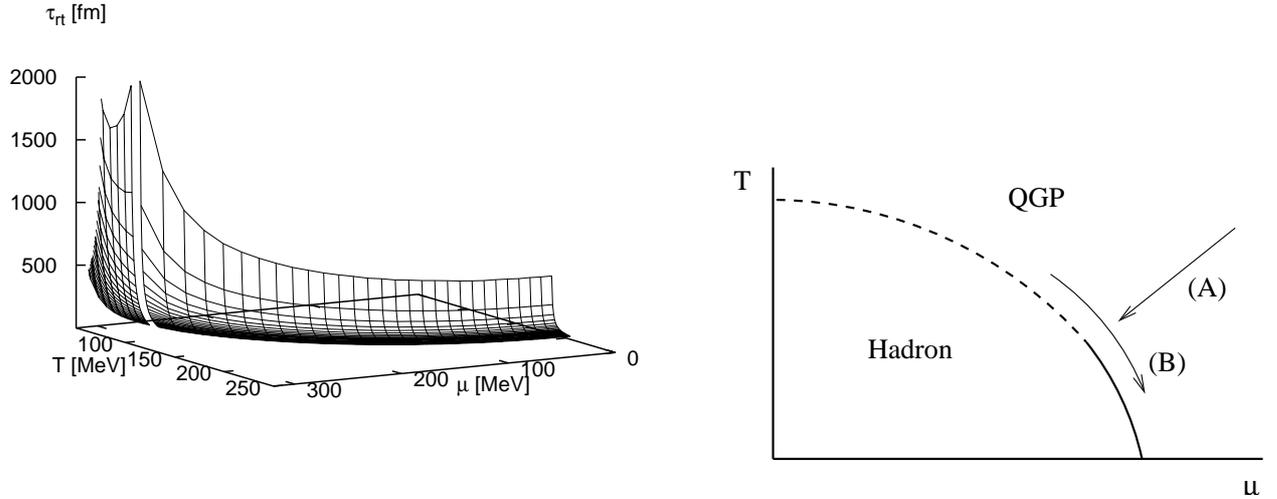

**FIGURE 2.** Temperature and chemical potential dependences of the relaxation time at higher temperature than 50 MeV (Left) and the schematic figure for two paths of temperature and chemical potential variations(A) and (B) (Right).

Furthermore, the enhancement around the tricritical point is sensitive for the variation along the critical line. When we change the temperature and the chemical potential along the direction perpendicular to the critical line (See (A) in the right panel of Fig.2), the enhancement is very small. The enhancement is remarkable along the critical line((B) of the right panel in Fig.2).